\pdfoutput=1
\documentclass{nature}

\usepackage{hyperref,graphicx,color}
\usepackage[labelformat=empty]{subfig}
\usepackage[font=singlespacing]{caption}

\newcommand{\denovo}[0]{\emph{de novo} }

\title{Bayesian sequence assembly and assessment by Markov Chain Monte Carlo sampling}
\author{Mark Howison\,$^{1}$, Felipe Zapata\,$^2$, Erika J. Edwards\,$^2$, and Casey W. Dunn\,$^{2}$}

\begin{document}
\maketitle

\begin{affiliations}
\item Center for Computation and Visualization, Brown University, Providence, RI, USA
\item Department of Ecology and Evolutionary Biology, Brown University, Providence, RI, USA
\end{affiliations}

\begin{abstract}
Most genome assemblers construct point estimates, choosing a genome sequence from among many alternative hypotheses that are supported by the data. We present a Markov Chain Monte Carlo approach to sequence assembly that instead generates distributions of assembly hypotheses with posterior probabilities, providing an explicit statistical framework for evaluating alternative hypotheses and assessing assembly uncertainty. We implement this approach in a prototype assembler and illustrate its application to the bacteriophage $\Phi$X174.
\end{abstract}

Most current methods for \denovo genome assembly generate point estimates of the genome sequence without explicit statistical information about confidence in this particular estimate, or its support relative to other alternative assemblies supported by the same sequence data\cite{howison_2013}.  Recently, several probabilistic approaches have been proposed to quantify assembly certainty and address these limitations. The Computing Genome Assembly Likelihood (CGAL) tool approximates the likelihood of an assembly given the sequence reads and a generative model learned from the data\cite{rahman_cgal:_2013}; the Log Average Probability (LAP) tool approximates the likelihood under a similar model of sequence read generation\cite{ghodsi_2013}; and the Assembly Likelihood Evaluation (ALE) framework uses an empirical Bayesian approach to estimate the posterior probability of a particular assembly (or components of that assembly)\cite{clark_ale:_2013}. 

Here we present an alternative Bayesian approach to approximating posterior probabilities, using Markov Chain Monte Carlo (MCMC) sampling\cite{gilks_markov_1995} to generate an entire posterior distribution of assembly hypotheses. An MCMC framework for sequence assembly addresses many of the general problems of accommodating assembly uncertainty, and provides a probabilistic understanding of the support for an assembly as a whole, as well as for particular features of the assembly that the investigator has special interest in. For example, it may be that a particular attribute of an assembly has low confidence, but that the other assemblies that together account for 100\% of the alternatives are all still consistent with a broader-scale hypothesis, such as the order of a set of genes or existence of a specific regulatory element.

To demonstrate the feasibility of the MCMC approach, we have implemented a prototype assembler called Genome Assembly by Bayesian Inference (GABI, \url{https://bitbucket.org/mhowison/gabi}). Because our approach is computationally expensive, our current tests are limited to the small but well-studied genome of the $\Phi$X174 phage (which has NCBI Reference Sequence \texttt{NC\_001422.1}). Our input data consist of 2,000 read-pairs drawn from Illumina $\Phi$X174 sample data (\url{http://www.illumina.com/systems/miseq/scientific\_data.ilmn}, estimated mean insert size of 357bp). These 250bp reads, available as Supplementary Data 1, provide approximately 85$\times$ coverage of the $\Phi$X174 genome.

Conveniently, our approach has technical parallels and similar conceptual advantages as the advances made over the past 15 years in the application of Bayesian approaches to phylogenetic analysis\cite{holder_phylogeny_2003, Ronquist2003, Lartillot2009}. As in phylogenetics and other fields, designing an MCMC sampling strategy that achieves good mixing and convergence poses a variety of challenges and opportunities. For sequence assembly, the challenge is to design a mechanism for proposing new assemblies, a means of calculating the likelihood of a proposed assembly under an explicit model of read generation, and the specification of a prior probability distribution that describes some prior beliefs about the assembly before the sequence data were collected. To implement an MCMC sequence assembler, we made the following choices:

\begin{description}
\item[Proposal mechanism.]
We first build an assembly graph starting from a de Bruijn graph of the reads. Then we remove all tips (since the $\Phi$X174 genome is circular) and merge all unambiguous paths into single edges that are annotated by the sequence of merged $k$-mers. The resulting unresolved assembly graph (no longer de Bruijn) is a directed multi-graph that consists only of bubbles and is a minimal representation of the variants that can be inferred from the sequenced data. The concatenated edge sequences for a particular path through this graph gives a possible assembly sequence. 

A particular assembly hypothesis is represented as a boolean vector, whose values indicate which of the enumerated edges in the assembly graph are active (Supplementary Fig. 1). We start with a random vector and propose new assemblies by toggling a value at random in the current vector. However, when toggling, we enforce the constraint that an active path cannot take multiple branches through the same bubble, since this would not spell a contiguous sequence. Instead, turning on an edge in an alternate branch turns on the \emph{entire} alternate branch and turns off the existing path through the bubble (Supplementary Fig. 2). In this way, we can propose alternate paths through a complex bubble in one move of the sampler, without having to split the path and wait for the alternate branch to randomly turn on.

In our prototype implementation, we are only considering paths that visit an edge at most once. In the general case, however, the proposal mechanism should be extended to accommodate paths that revisit edges, in order to resolve repeat regions of a genome. 

\item[Likelihood.]
Because of the constraints described above, the active paths in every proposed assembly can be output as a FASTA file of concatenated edge sequences. We run LAP directly on this FASTA file to estimate the proposed assembly's likelihood.

\item[Priors.]
An uninformative (e.g., flat) prior distribution could be used if nothing is known about the genome to be assembled. Since $\Phi$X174 is well known, and it has previously been established that the genome is 5,386 bp and consists of one circular chromosome, we construct a prior probability distribution as the product of two gamma distributions, one for the sum of contig lengths (centered at 5,386) and the other for the number of contigs in the assembly (centered at 1). In other cases, external information about repeat structure or gene synteny could also be incorporated as priors.

\end{description}

To evaluate the mixing and convergence of our MCMC sampler, we ran three independent chains and compared their traces and split frequencies, as is common practice for other applications of MCMC\cite{Nylander2008}. Our results show that our design achieves good mixing when using a simple Metropolis sampler\cite{Metropolis1953} without any burn-in or thinning (Fig. 1a). The cumulative frequencies of the individual edges have mostly stabilized after 8,000 accepted samples (Fig. 1b). Those with weak support were likely assembled from reads with sequencing errors. Other edges are more ambiguous, with a cumulative frequency that hovers at an intermediate value or varies across samples. Overall, the standard deviation in edge frequencies between the chains decreases with additional sampling, indicating that independent chains are converging to the same posterior distribution (Fig. 1c). The split frequencies among three independent chains are mostly correlated after the last sample (Fig. 1d). We also tested the sampler with flat priors, with no likelihood calculations (priors only), and with different choices of the scaling parameter for the gamma distributions (Supplementary Fig. 3). Our acceptance rate is 42.2\%, which is higher than the heuristic of 25\% that can be considered optimal for general Metropolis sampling\cite{Roberts1997}. Aggregated across all three chains, the mean compute time per sample was 1.6 seconds and total compute time was 25,324.5 CPU-hours.

GABI provides multiple ways for summarizing the results of an MCMC analysis and is built on top of BioLite, a light-weight bioinformatics framework with rich diagnostics and reporting capabilities\cite{howison_biolite_2012}. The assembly graph can be annotated with the approximated posterior probabilities (Fig. 2a), or a consensus assembly can be extracted, for instance a majority-rule consensus that shows all edges occurring with frequency $>50\%$ (Fig. 2b). The report includes a FASTA file for the majority-rule consensus, annotated with the posterior probabilities of its components. A detailed report created with the D3js data visualization toolkit\cite{Bostock2011} provides an interactive animation of the sampling for each MCMC chain. This report is provided for all chains as Supplementary Data 2 (and also at \url{http://ccv.brown.edu/mhowison/gabi-report}).

GABI includes a tool to assign posterior probabilities to features of external assemblies that correspond to features in its own assembly graph. This provides an explicit and unified statistical framework for comparison of assemblies produced by multiple methods and software tools. Here, we compare  NCBI Reference Sequence \texttt{NC\_001422.1} and \denovo point-estimate assemblies generated by the String Graph Assembler\cite{Simpson2012} and Velvet\cite{Zerbino2008}. We require exact matches to identify corresponding features, and this conservative strategy means that there is not an exact correspondence between the NCBI reference sequence and the GABI graph (Fig. 2e). For this simple assembly problem, there is nearly universal agreement among the assemblies: both the majority rule consensus (Fig. 2b) and NCBI reference sequence (Fig. 2e) are proper subsets of the two \emph{de novo} point-estimate assemblies (Fig. 2c-d). The one notable difference is that the SGA assembly contains two contigs that choose alternate paths in one of the bubbles, and these alternate paths have similar posterior probabilities (yellow highlight in Fig. 2).

A challenge to scaling MCMC assembly is that the combinatorial complexity of larger assembly graphs could become prohibitive for full \denovo Bayesian assembly of large genomes. There are several ways to address this, such as applying more sophisticated sampling methods like Metropolis-coupled MCMC\cite{Geyer1991} or bridging states\cite{Lin2012} that can explore larger combinatorial spaces more efficiently; constraining the assembly graph to focus on particular assembly hypotheses instead of attempting full \denovo assembly; or pruning the assembly graph using additional data from restriction site mappings\cite{lam_genome_2012}. Another promising application is transcriptome assembly, since the assembly graphs for individual transcripts should be less complex and can be sampled independently.

Like existing assemblers, GABI can be used to assemble a point estimate of a genome, but unlike most other assemblers, the resulting assembly will have been chosen according to explicit statistical criteria (posterior probability) and will have associated information on the confidence in various aspects of its sequence and structure. In addition, MCMC assembly provides new opportunities for investigators who are interested not only in the certainty of a particular inference (e.g. an ALE\cite{clark_ale:_2013} estimate of posterior probability for a given assembly), but in the many alternative hypotheses that are also supported by the data. The MCMC approach addresses many of the general problems of summarizing assembly uncertainty and will allow assembly uncertainty to be propagated to downstream analyses\cite{howison_2013}.

\begin{methods}
The results presented were generated with GABI version 1.1.2 and are recomputable using included scripts (\url{https://bitbucket.org/mhowison/gabi/src/master/phix-test}). Here, we provide brief comments on the technical details.

\subsection{Compute resources.}

All tests were run at the Center for Computation and Visualization, Brown University, on IBM iDataPlex nodes, each equipped with 16-core, dual-socket Intel Xeon E5-2670 (2.6Ghz) processors and 64GB of memory. CPU-hours were calculated as total walltime across all nodes, multiplied by 16 CPUs per node.

\subsection{Subset of $\Phi$X174 sample data.}

We started with 8.36 Gb of paired-end reads of length 250 bp from Illumina $\Phi$X174 sample data (\url{ftp://webdata:webdata@ussd-ftp.illumina.com/Data/SequencingRuns/PhiX/PhiX_S1_L001_R1_001.fastq.gz} and \url{ftp://webdata:webdata@ussd-ftp.illumina.com/Data/SequencingRuns/PhiX/PhiX_S1_L001_R2_001.fastq.gz}, accessed April 26, 2013). Using the \texttt{illumina\_filter} tool from BioLite, we chose the first 2,000 read pairs that did not contain known Illumina adapter sequences and that had mean quality score greater than 37 (Phred-33 scoring). This procedure is available in the script \url{https://bitbucket.org/mhowison/gabi/src/master/phix-test/00-subset-data.sh}.

\subsection{SGA and Velvet assemblies.}

To assemble the 2,000 read $\Phi$X174 subset with SGA, we followed the example provided for an assembly of E. coli from similar MiSeq reads (\url{https://github.com/jts/sga/blob/master/src/examples/sga-ecoli-miseq.sh}). To assemble with Velvet, we used the included VelvetOptimiser tool to sweep coverage and cutoff parameters for a $k$-mer size of 99. The estimated mean and standard deviation of the insert size reported earlier were obtained from VelvetOptimiser's output. This procedure is available in the script \url{https://bitbucket.org/mhowison/gabi/src/master/phix-test/01-assemble.sh}

\subsection{De Bruijn graph reduction.}

First, we recursively and completely trimmed all tips (edges with an incidence of one), because our target $\Phi$X174 genome is circular (for a linear target, a softer tip trimming threshold would be more appropriate). Next, we merged all simple paths through the graph, similar to the process of reducing an \emph{overlap graph} to a \emph{string graph}\cite{myers_fragment_2005}. We annotated the new edge with the accumulated overlap of the merged nodes' $k$-mers, which accumulates one additional nucleotide for each merged edge. Finally, we split the graph into weakly connected components and choose the largest one. A cyclical path of edges through the graph spells a contig by concatenating the annotations on the edges. For a linear path, the contig starts with the $k$-mer at the initial node, followed by the concatenated edge annotations.

\subsection{Odds ratio.}

To determine whether to accept a proposed assembly hypothesis, we calculate the odds ratio\cite{Metropolis1953} of the posterior probabilities of the old assembly, $P(H|D)$, and the new perturbed assembly, $P(H^*|D)$, as:

\begin{equation}
R=\frac{P(H^*|D)}{P(H|D)} = \frac{ \frac{P(D|H^*)P(H^*)}{P(D)} } { \frac{P(D|H)P(H)}{P(D)} } = \frac{P(D|H^*)P(H^*)}{P(D|H)P(H)}
\end{equation}

If this ratio is greater than 1, the new assembly is automatically accepted. If it is less than 1, it is accepted with probability R. The process is then repeated until there is a stationary distribution of assemblies in the sample, which occurs when the frequency of assembly attributes does not change with additional sampling.
\end{methods}

\bibliographystyle{naturemag}
\bibliography{gabi}

\begin{addendum}
\item This work was supported by the National Science Foundation through the Alan T. Waterman Award to C.W.D. and award DEB-1026611 to E.J.E., and through additional support from the Brown Division of Biology and Medicine to E.J.E. This research was conducted using computational resources and services at the Center for Computation and Visualization, Brown University.
 \item[Author Contributions] M.H., F.Z., C.W.D. conceived the MCMC approach and prepared the manuscript. M.H. implemented the software, collected all results, and developed the methods with feedback from F.Z., C.W.D., E.J.E.
 \item[Competing Interests] The authors declare that they have no
competing financial interests.
 \item[Correspondence] Correspondence and requests for materials
should be addressed to M.H.~(email: mhowison@brown.edu).
\end{addendum}

\linespread{1}
\renewcommand{\figurename}{\bf Figure}

\begin{figure}[p!]
\centering
\includegraphics[width=\textwidth]{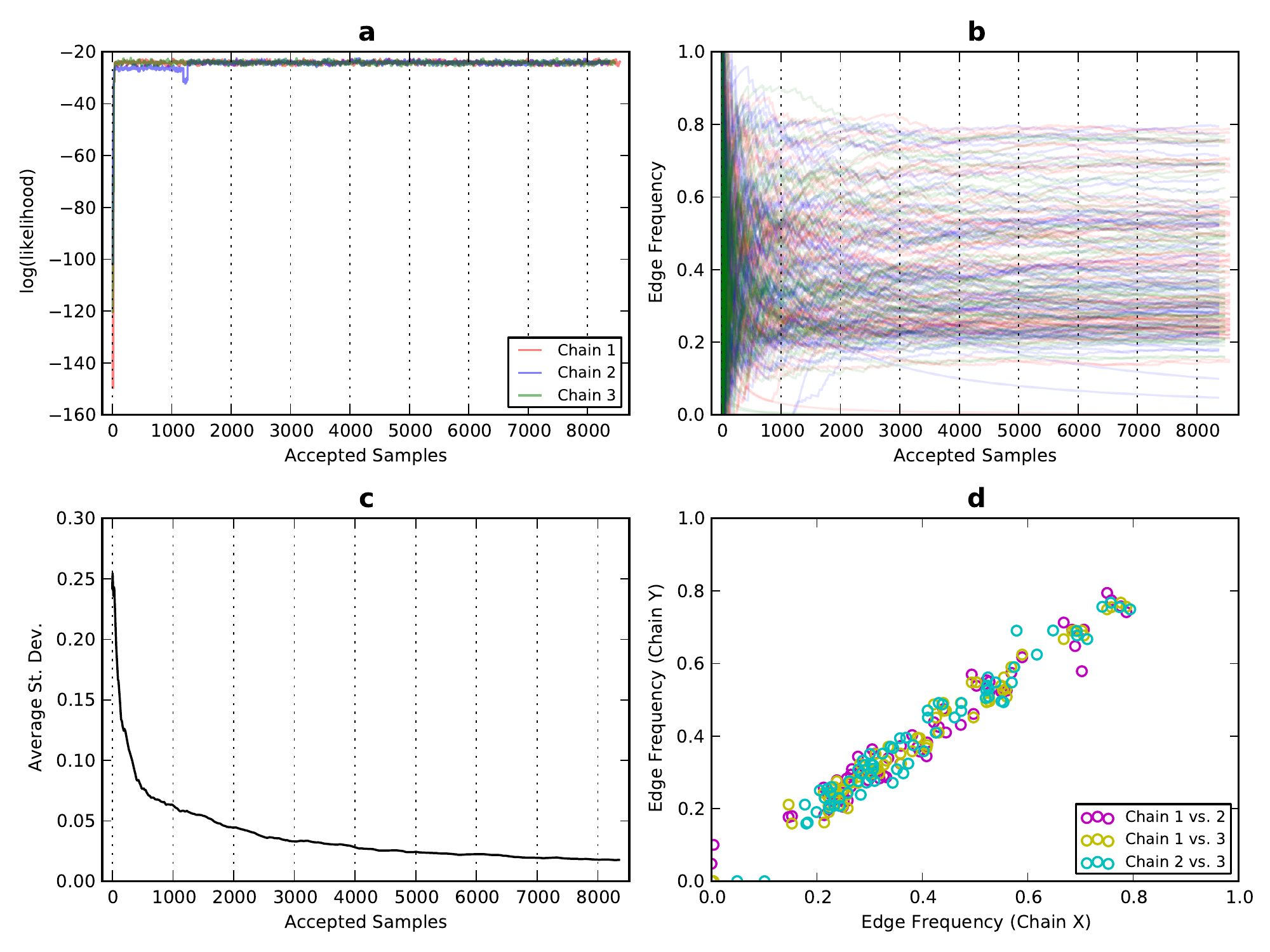}
\caption{Evidence of good mixing and convergence of two independent MCMC assembly chains. \textbf{(a)} Early in the sampling, the log(likelihood) reaches a stationary distribution with random noise, indicating good mixing of the chains. \textbf{(b)} The cumulative frequencies for graph edges shows most have reached a stationary value. \textbf{(c)} The average standard deviation (between the three chains) of the cumulative edge frequencies approaches zero. \textbf{(d)} Bivariate plot of the edge frequencies between each pair of chains after the final sample. Both \textbf{(c)} and \textbf{(d)} indicate convergence.}
\end{figure}

\begin{figure}[p!]
\centering
\subfloat[\textbf{a}]{\includegraphics[width=0.5\textwidth]{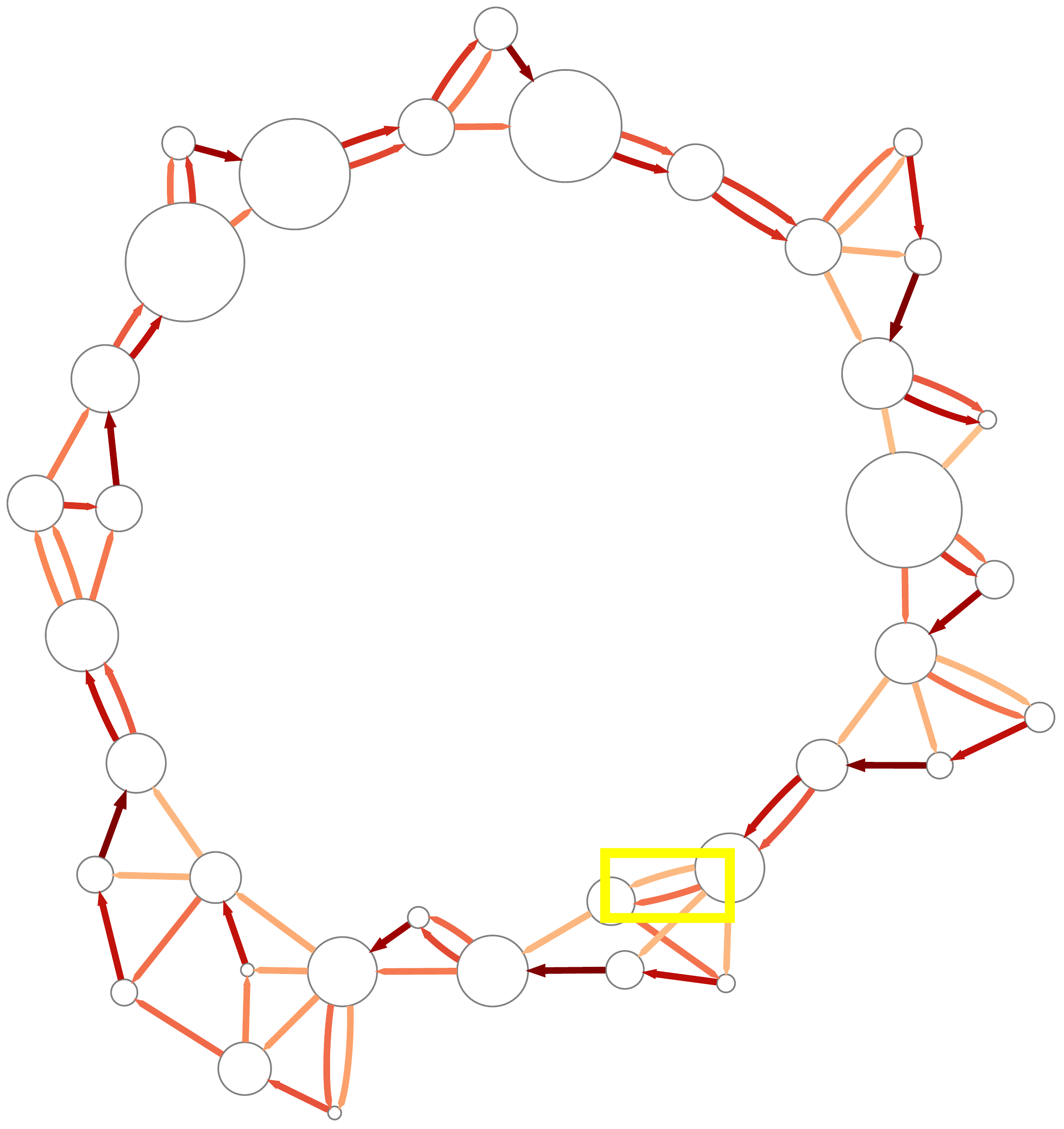}}
\subfloat[\textbf{b}]{\includegraphics[width=0.5\textwidth]{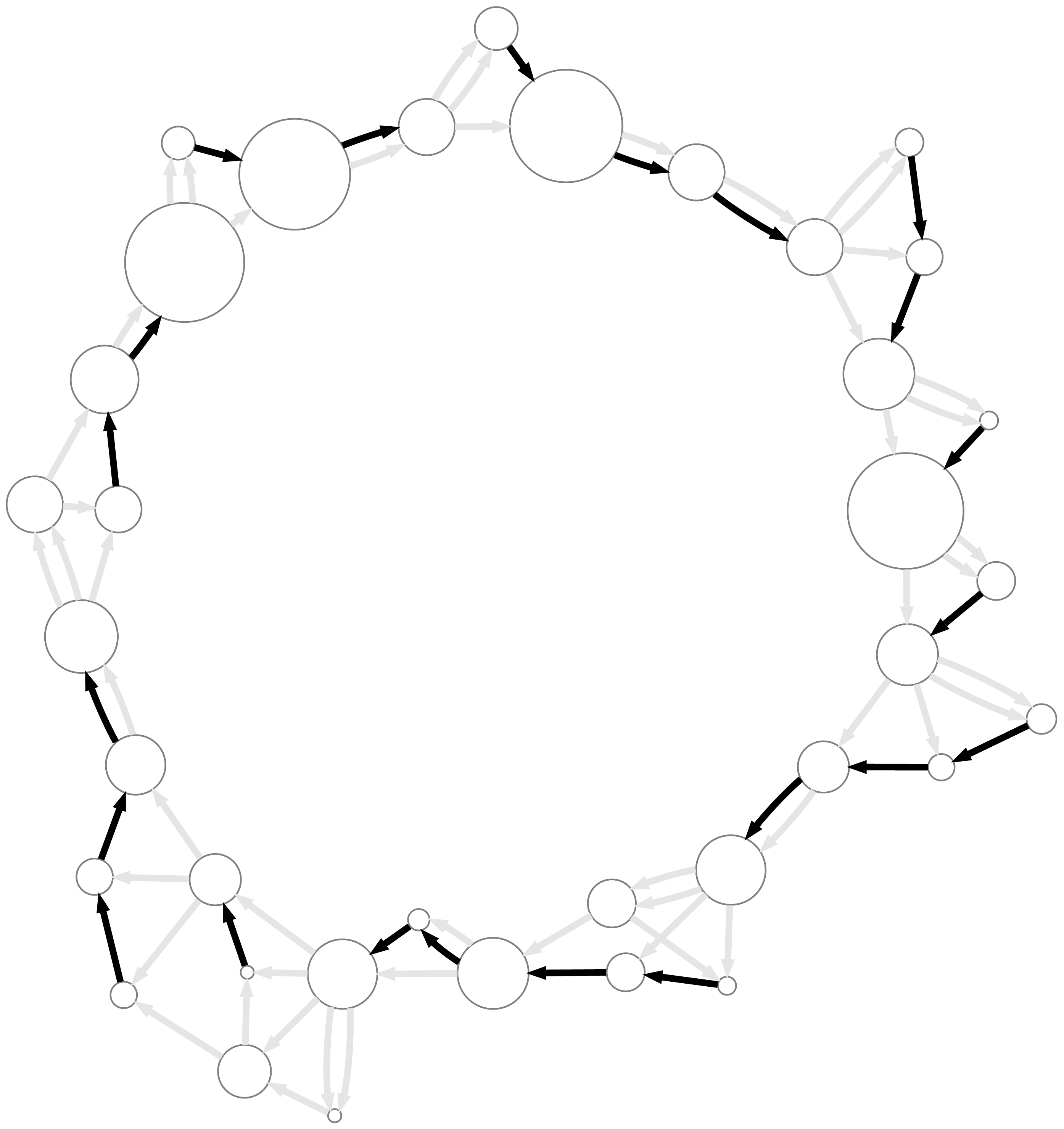}}
\newline
\subfloat[\textbf{c}]{\includegraphics[width=0.33\textwidth]{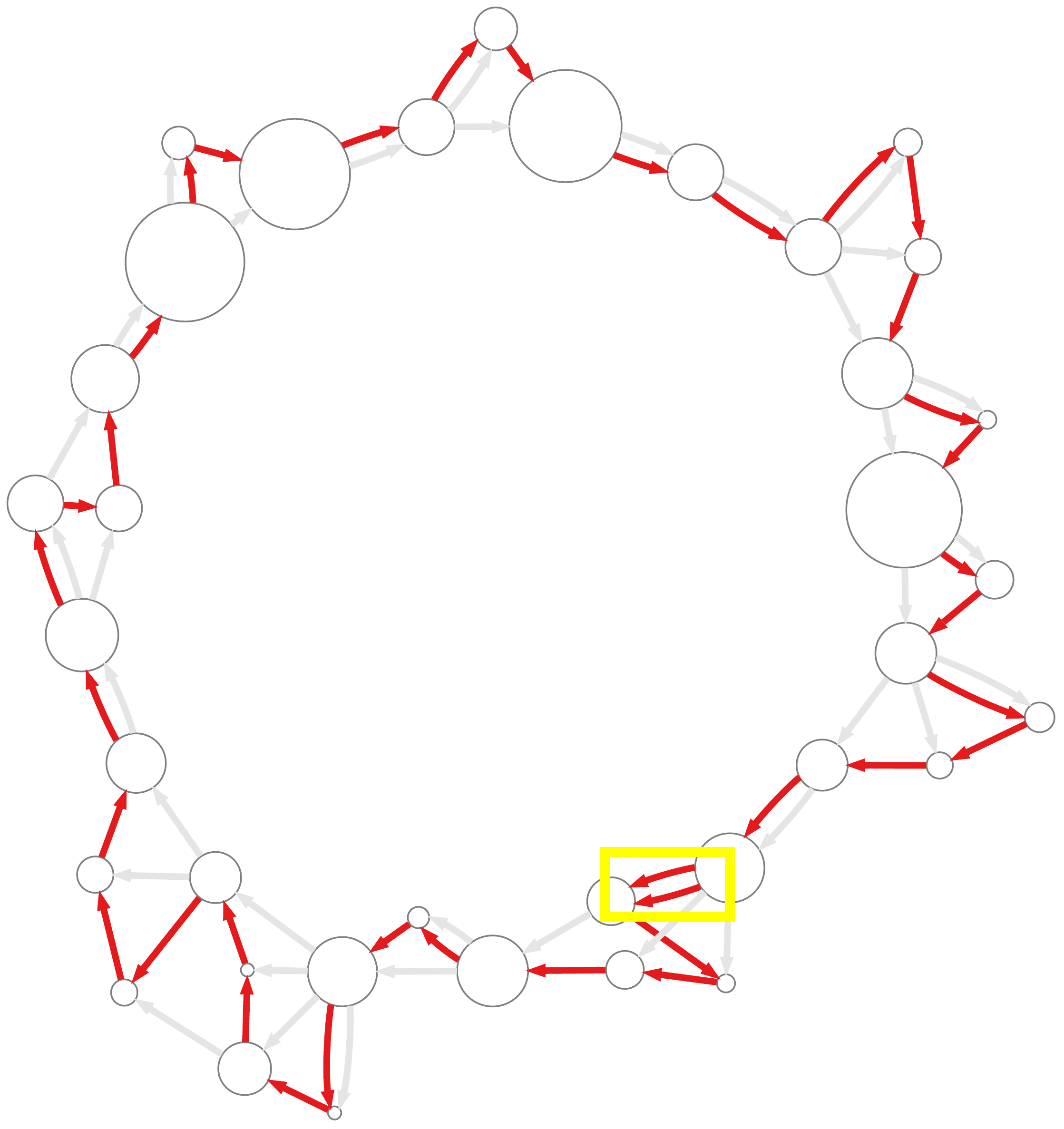}}
\subfloat[\textbf{d}]{\includegraphics[width=0.33\textwidth]{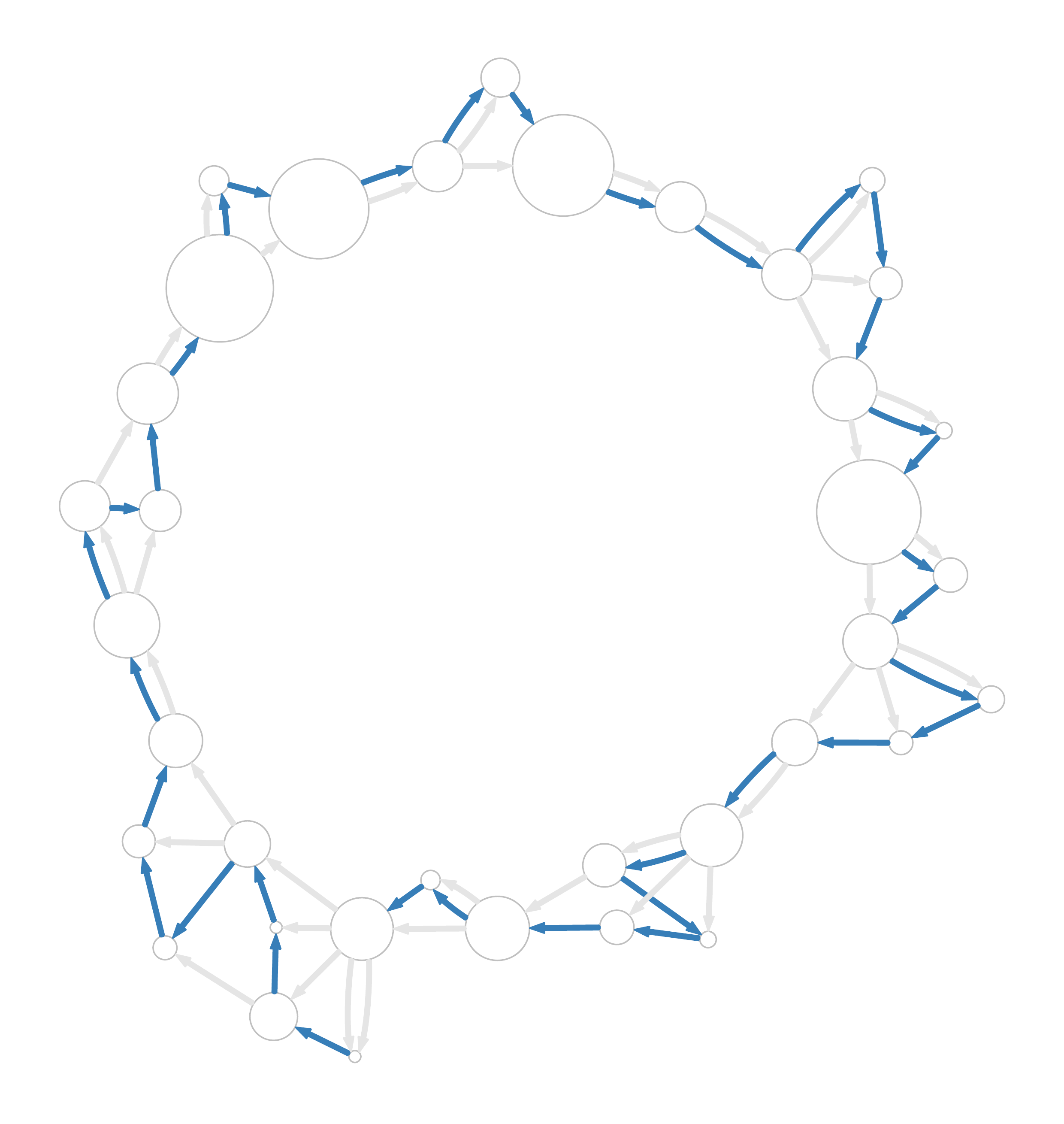}}
\subfloat[\textbf{e}]{\includegraphics[width=0.33\textwidth]{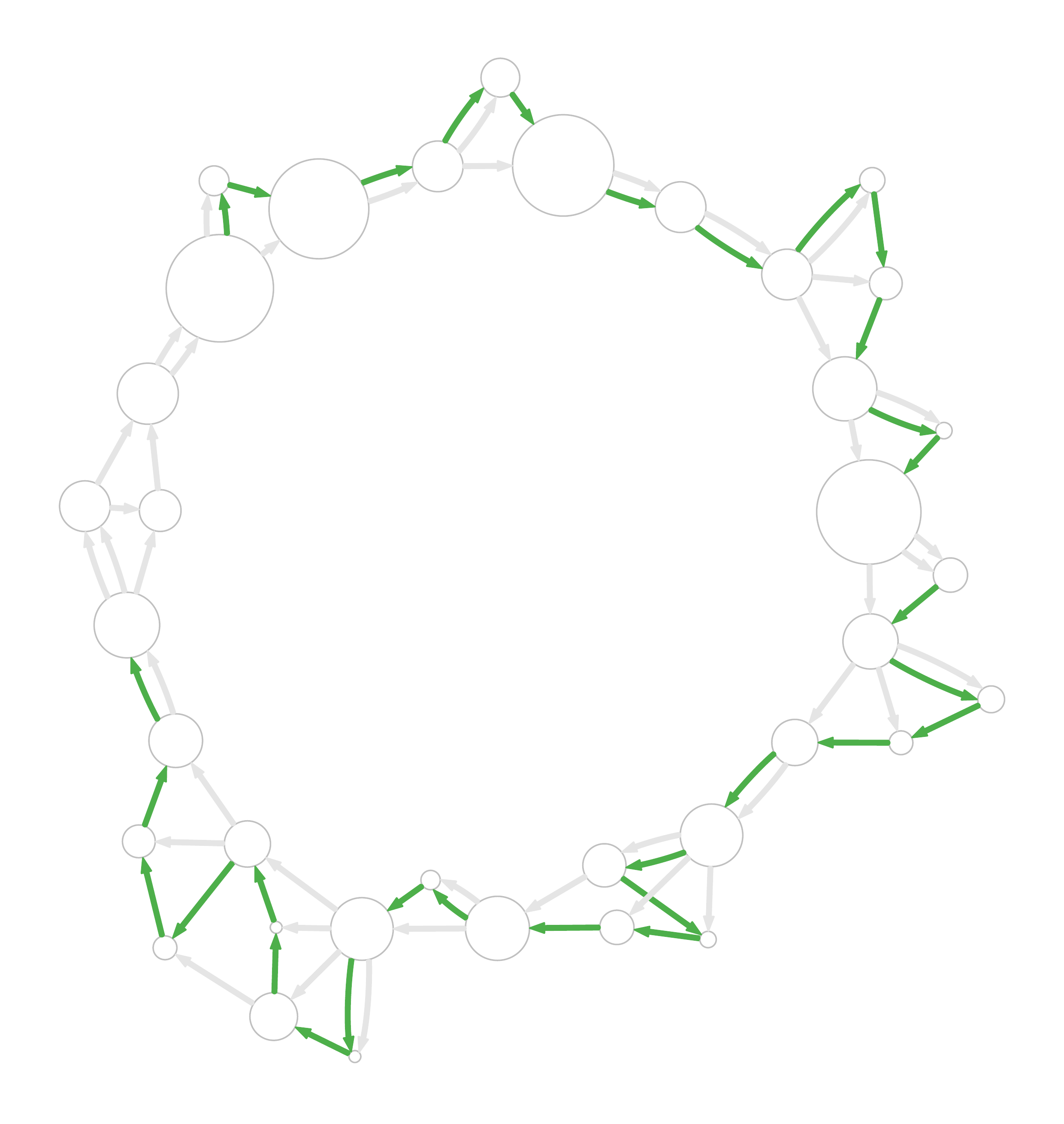}}
\caption{Graphical summaries of the $\Phi$X174 genome assemblies. Each edge is a sequence segment, and alternative paths through the graph represent alternative assemblies. Node size is proportional to the length of sequences emanating from the node. \textbf{(a)} Posterior probabilities for each edge (the darker the edge, the higher the posterior). \textbf{(b-e)}­ Edges from the GABI assembly graph that are found exactly in the majority­-rule consensus GABI assembly \textbf{(b)}, SGA assembly \textbf{(c)}, Velvet assembly \textbf{(d)}, and NCBI reference sequence \textbf{(e)}.}
\end{figure}

\setcounter{figure}{0}
\renewcommand{\figurename}{\bf Supplementary Figure}

\begin{figure}[p!]
\centering
\includegraphics[width=\textwidth]{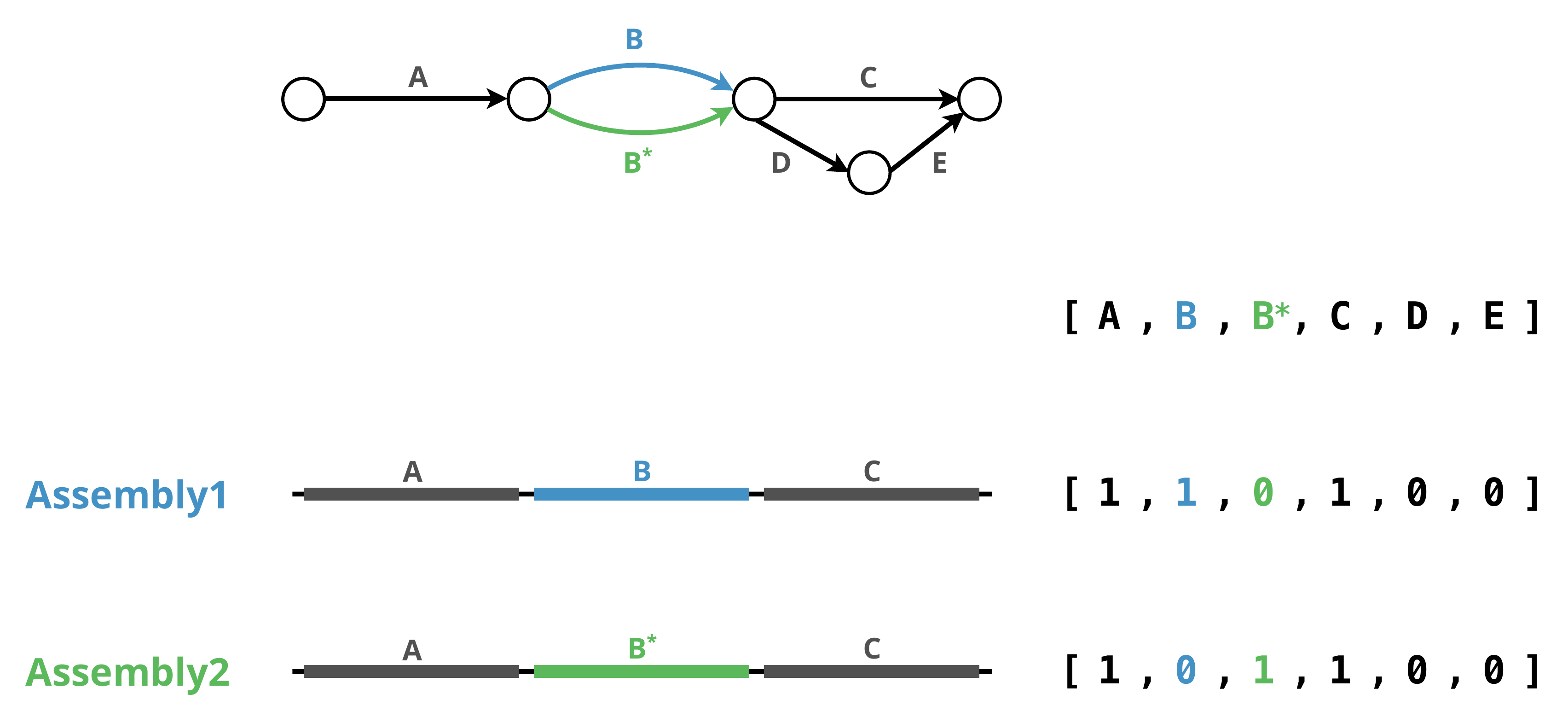}
\caption{
Contiguous paths through the assembly graph correspond to proposed assemblies, and are represented by a boolean vector indicating which edges in the graph are active.}
\end{figure}

\begin{figure}[p!]
\centering
\subfloat[\textbf{a}]{\includegraphics[width=0.5\textwidth]{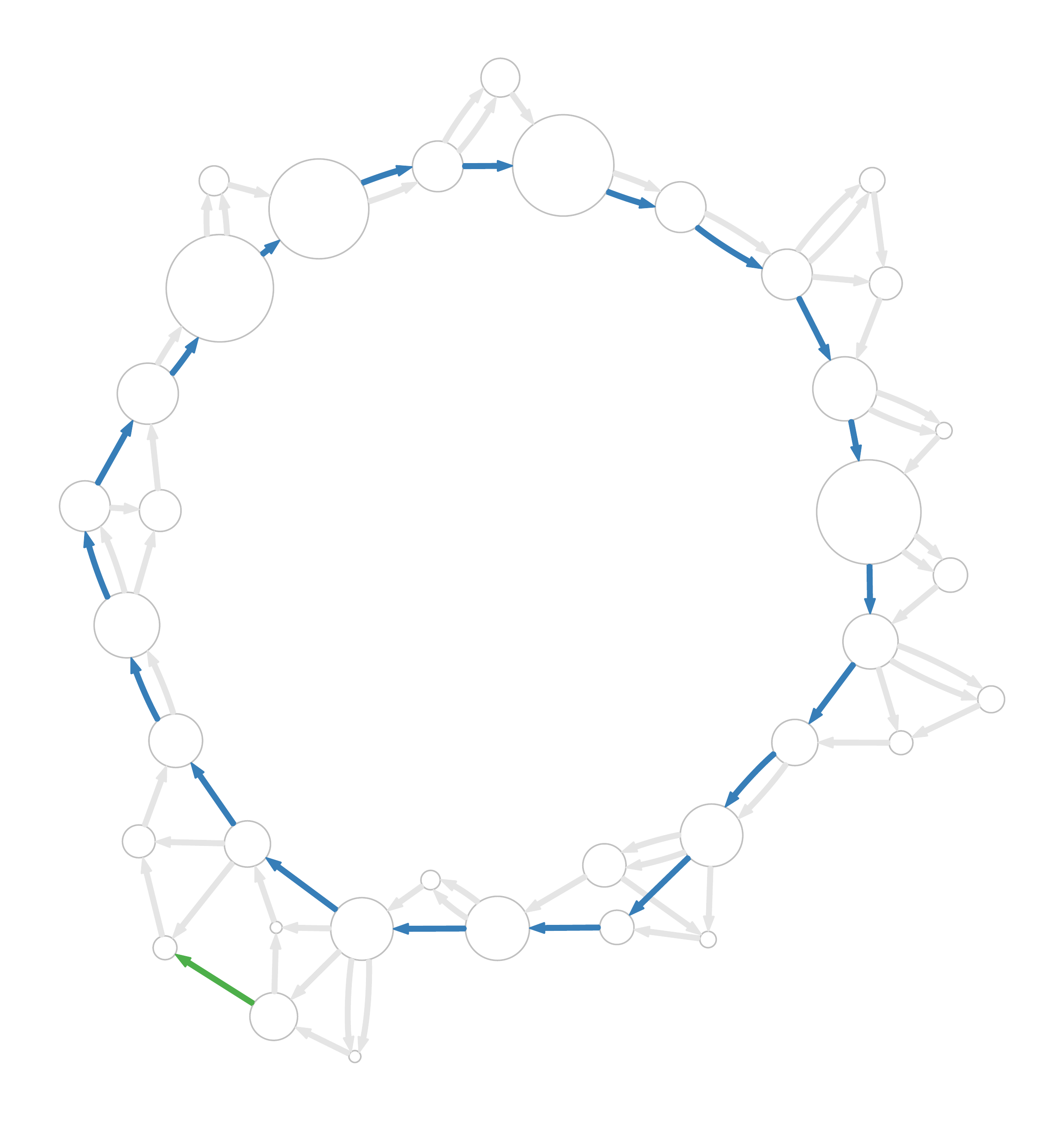}}
\subfloat[\textbf{b}]{\includegraphics[width=0.5\textwidth]{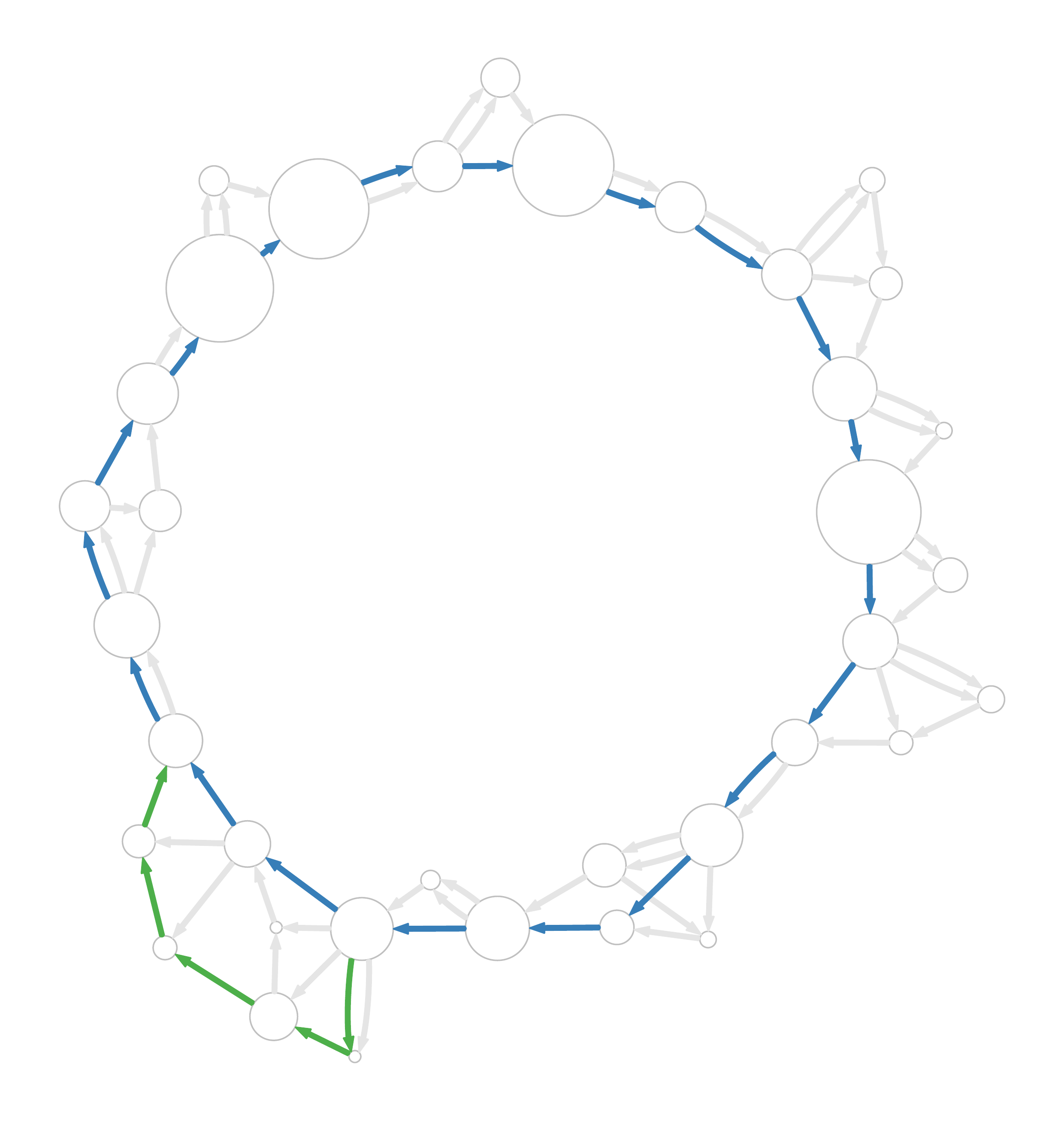}}
\newline
\subfloat[\textbf{c}]{\includegraphics[width=0.5\textwidth]{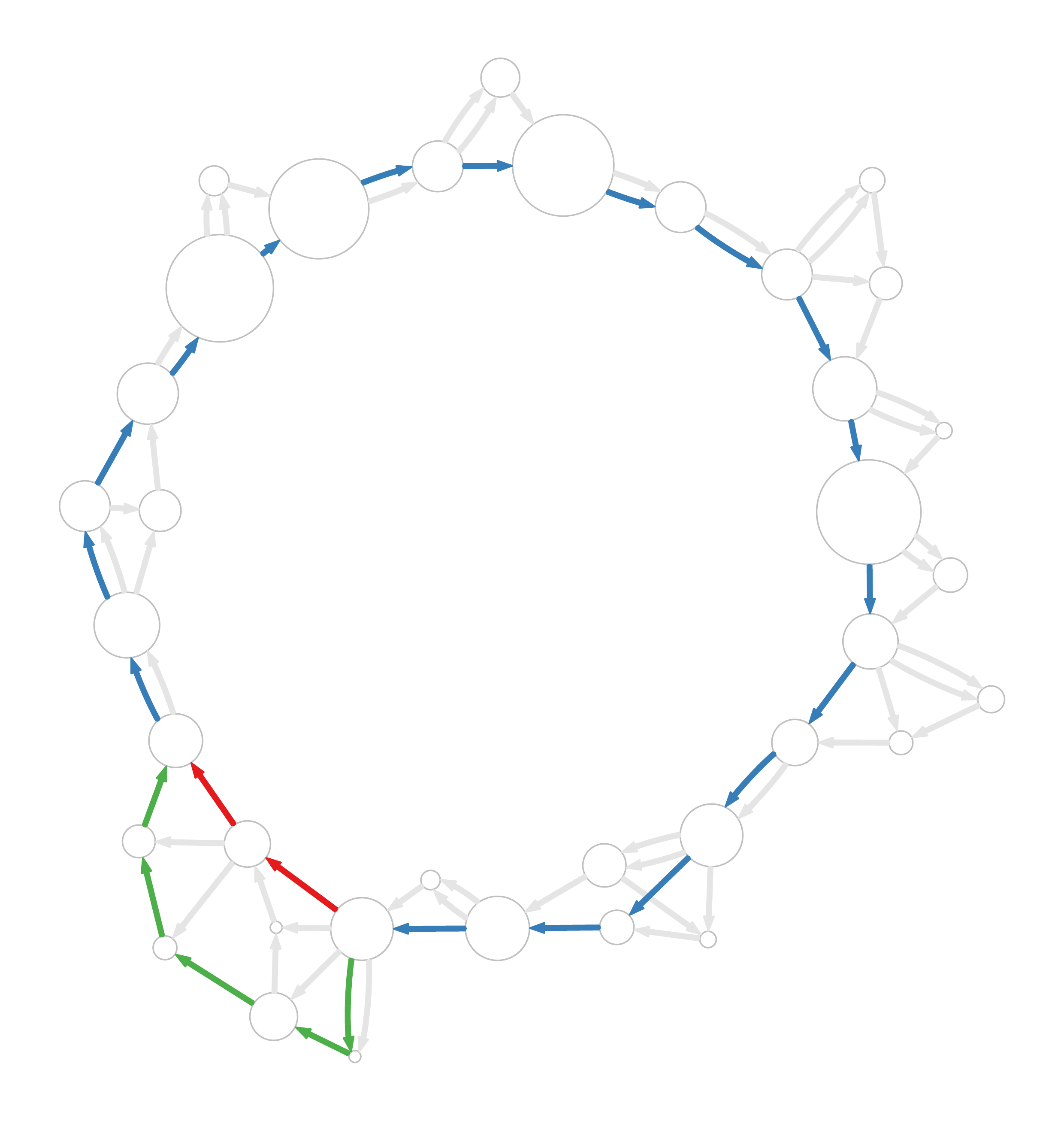}}
\subfloat[\textbf{d}]{\includegraphics[width=0.5\textwidth]{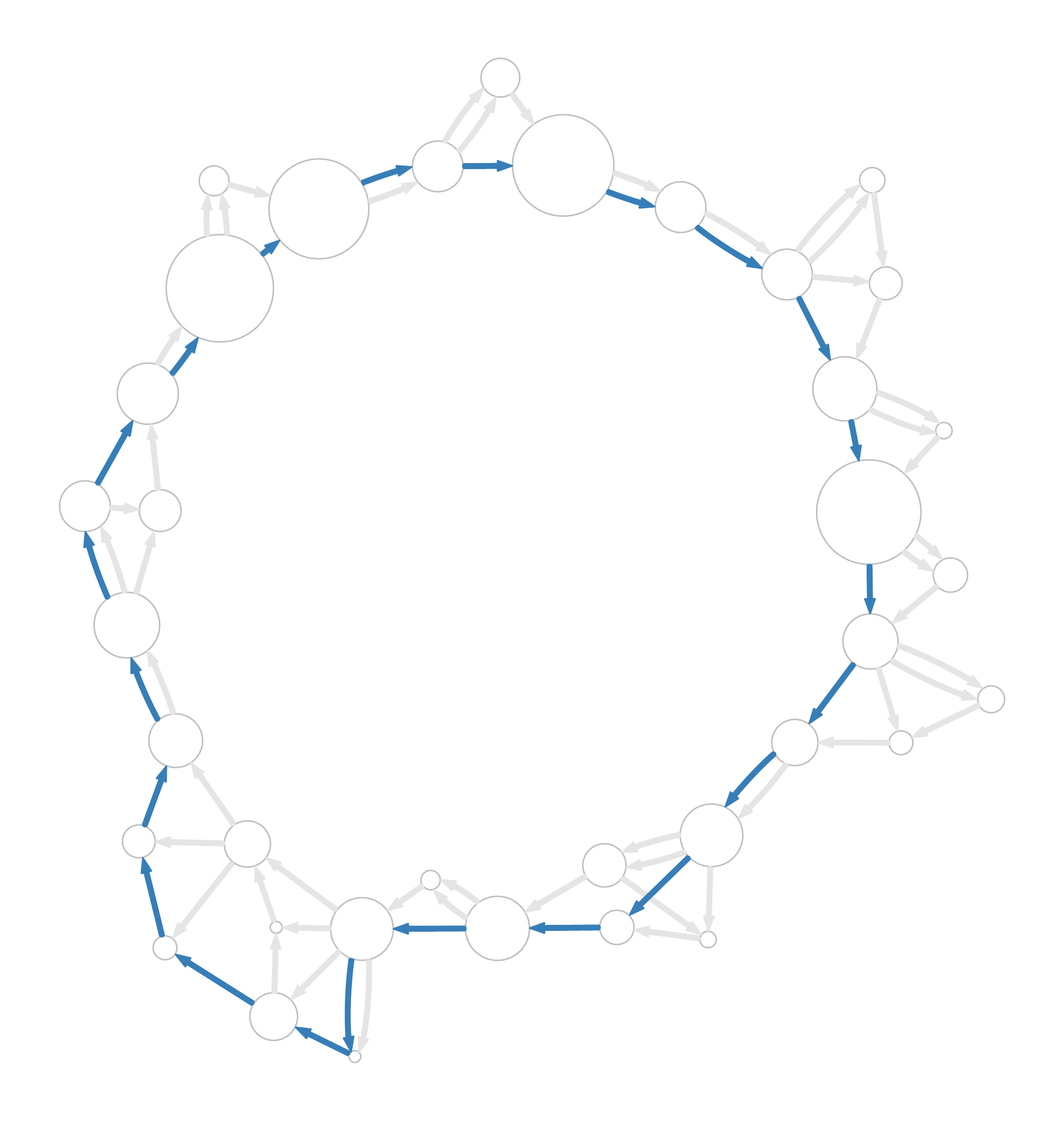}}
\caption{
\textbf{(a)} Starting from an existing (blue) assembly, the proposal mechanism randomly chooses a new (green) edge. \textbf{(b)} If the edge is not already active, it is extended with a random walk until it meets the existing assembly. \textbf{(c)} This defines a (red) branch in the existing assembly, \textbf{(d)} which is then removed to yield a new (blue) assembly.}
\end{figure}

\begin{figure}[p!]
\centering
\includegraphics[width=0.8\textwidth]{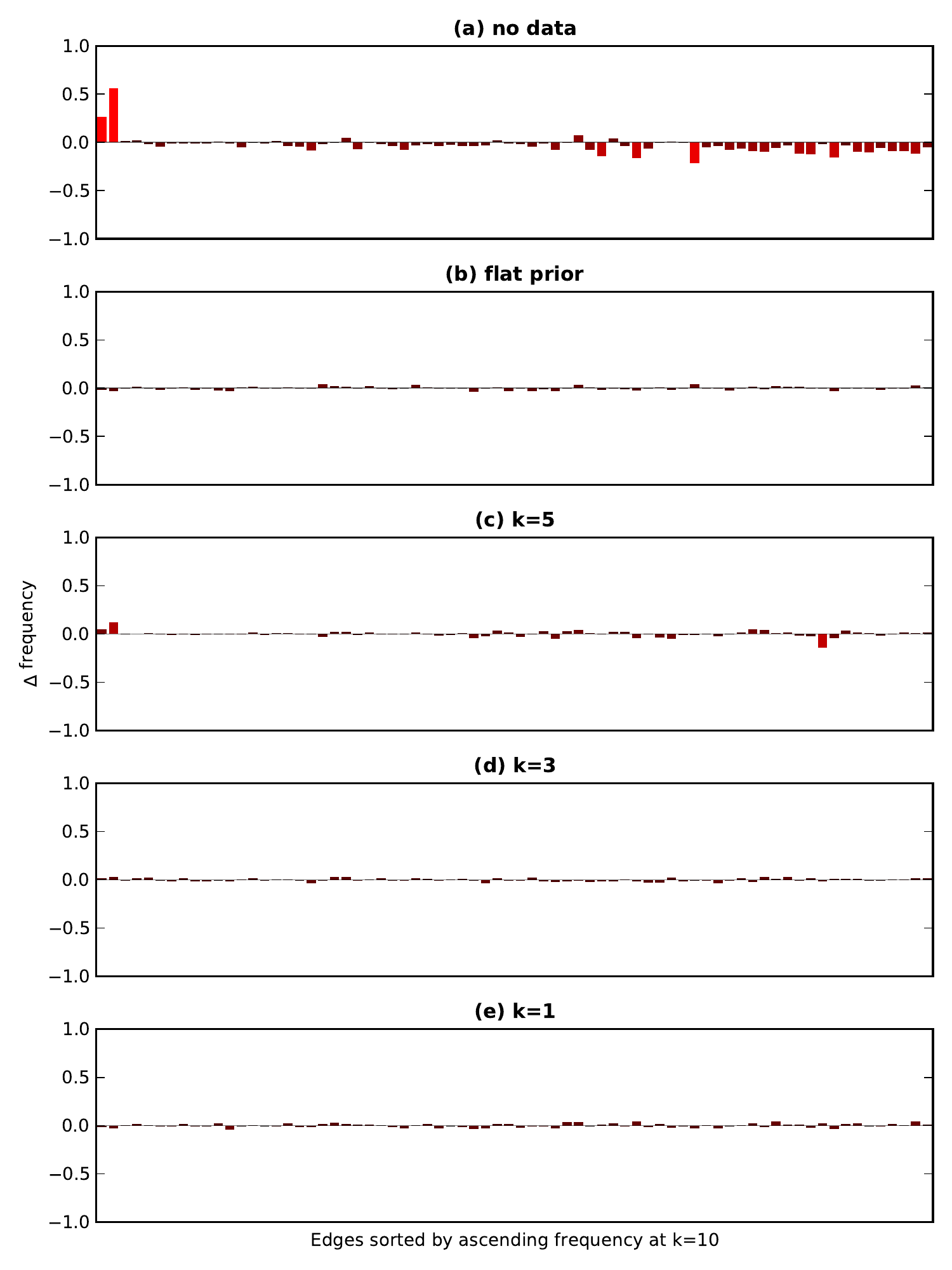}
\caption{The proposed MCMC sampler is driven by the likelihood calculations and is robust to changes in priors. Running the sampler with only the prior probabilities and no likelihood calculations \textbf{(a)} causes many of the edge frequencies in the posterior distribution to diverge. But running the sampler with flat priors \textbf{(b)} or with different values of the scaling parameter $k$ for the priors on genome length and number of contigs \textbf{(c-e)} causes only minor variation in the edge frequencies.}
\end{figure}
\end{document}